\documentclass[conference]{IEEEtran}
\IEEEoverridecommandlockouts
\usepackage[pdftex]{graphicx}
\graphicspath{{figs/}{/figs/}}
\DeclareGraphicsExtensions{.pdf,.jpeg,.png}

\usepackage{url}
\usepackage[dvipsnames]{xcolor}
\usepackage[para]{threeparttable}
\usepackage{array}

\usepackage[utf8]{inputenc}
\usepackage[T1]{fontenc}
\usepackage{bm} % for bold math symbols (incl. Greek letters) with \bm{}
\usepackage{amsmath,times,booktabs, tabularx, hyperref, subcaption, multirow}
\usepackage{amssymb, amsfonts}
\usepackage{microtype}

\usepackage[%
        backend=biber,                          %... Set ref. backend to biber
        style=ieee,                             %... Set final ref. style to IEEE
    ]{biblatex}
\DeclareSourcemap{
  \maps[datatype=bibtex]{
    \map{
      \step[fieldsource=publisher, match=arXiv, final]
      \step[fieldset=howpublished, fieldvalue={arXiv Preprint}]
    }
  }
}

\usepackage{algorithmic}
\usepackage{textcomp}

\def\BibTeX{{\rm B\kern-.05em{\sc i\kern-.025em b}\kern-.08em
    T\kern-.1667em\lower.7ex\hbox{E}\kern-.125emX}}

\usepackage[capitalise, nameinlink]{cleveref}
\usepackage{xcolor}

\usepackage{siunitx}
\usepackage[automake, nomain, acronym, xindy, shortcuts=ac]{glossaries-extra}
\glsdisablehyper
\makeglossaries
\setabbreviationstyle[acronym]{long-short} %... set long then short abbrev
\glssetcategoryattribute{acronym}{nohyperfirst}{true}
\newacronym{ae}{AE}{autoencoder}
\newacronym{ann}{ANN}{artificial neural network}
\newacronym{aru}{ARU}{autonomous recording unit}
\newacronym{ast}{AST}{Audio Spectrogram Transformer}
\newacronym{auc}{AUC}{area under the ROC curve}
\newacronym{atbfl}{ATBFL}{Acoustic Trends Blue Fin Library}
\newacronym{atp}{ATP}{Acoustic Trends Project}
\newacronym{aucroc}{AUC-ROC}{area under the receiver operating characteristic}
\newacronym{ap}{AP}{average precision}
\newacronym{bn}{BN}{bottleneck network}
\newacronym{beans}{BEANS}{Benchmark of Animal Sounds}
\newacronym{bilstm}{Bi-LSTM}{bidirectional long short-term memory}
\newacronym{cnn}{CNN}{convolutional neural network}
\newacronym{crnn}{CRNN}{convolutional recurrent neural network}
\newacronym{dct}{DCT}{Discrete Cosine Transform}
\newacronym{dnn}{DNN}{deep neural network}
\newacronym{dl}{DL}{deep learning}
\newacronym{dcase}{DCASE}{Detection and Classification of Acoustic Scenes and Events}
\newacronym{em}{EM}{expectation-maximisation}
\newacronym{fsl}{FSL}{few-shot learning}
\newacronym{fp}{FP}{false positive}
\newacronym{fn}{FN}{false negative}
\newacronym{fb}{FB}{False Bay}
\newacronym{fssed}{FSSED}{few-shot sound event detection}
\newacronym{gmm}{GMM}{Gaussian Mixture Model}
\newacronym{gru}{GRU}{gated recurrent unit}
\newacronym{hmm}{HMM}{Hidden Markov Model}
\newacronym{iou}{IoU}{Intersection over Union}
\newacronym{iwc-sorp}{IWC-SORP}{International Whaling Commission's Southern Ocean Research Partnership}
\newacronym{ltsa}{LTSA}{long-term spectral averages}
\newacronym{lfdcs}{LFDCS}{Low Frequency Detection and Classification System}
\newacronym{lfcc}{LFCC}{linear frequency cepstral coefficients}
\newacronym{lstm}{LSTM}{long short-term memory}
\newacronym{mfcc}{MFCC}{mel frequency cepstral coefficients}
\newacronym{mlp}{MLP}{multilayer perceptron}
\newacronym{map}{mAP}{mean average precision}
\newacronym{noaa}{NOAA}{National Oceanic and Atmospheric Administration}
\newacronym{ncei}{NCEI}{National Centers for Environmental Information}
\newacronym{nn}{NN}{neural network}
\newacronym{pam}{PAM}{Passive Acoustic Monitoring}
\newacronym{pcen}{PCEN}{per-channel energy normalization}
\newacronym{protonet}{ProtoNet}{prototypical network}
\newacronym{pipan}{PIPAN}{Pacific Islands Passive Acoustic Network}
\newacronym{rnn}{RNN}{recurrent neural network}
\newacronym{resnet}{ResNet}{residual network}
\newacronym{relnet}{RelNet}{relation network}
\newacronym{relu}{ReLU}{rectified linear unit}
\newacronym{roc}{ROC}{Receiver Operating Characteristic}
\newacronym{sed}{SED}{sound event detection}
\newacronym{sanctsound}{Sanctsound}{Sanctuary Soundscape Monitoring Project}
\newacronym{stft}{STFT}{short-time Fourier transform}
\newacronym{svm}{SVM}{support vector machine}
\newacronym{sisnr}{SISNR}{scale-invariant signal-to-noise ratio}
\newacronym{snr}{SNR}{signal-to-noise ratio}
\newacronym{scl}{SCL}{supervised contrastive learning}
\newacronym{tdnn}{TDNN}{Time Delay Neural Network}
\newacronym{tim}{TIM}{Transductive Information Maximisation}
\newacronym{tp}{TP}{true positive}
\newacronym{tn}{TN}{true negative}
\newacronym{vad}{VAD}{voice activity detection}
\newacronym{wmmsd}{WMMSD}{Watkins Marine Mammal Sound Database}
\newacronym{zcr}{ZCR}{zero crossing rate}

\newglossaryentry{wadbpn}{
	name={WhaleVAD-BPN},
	description={},
}

\newglossaryentry{wad}{
	name={WhaleVAD},
	description={},
}

\newglossaryentry{perch}{
	name={Perch},
	description={},
}

\newcommand{\emailaddr}[1]{{\small\normalfont\texttt{#1}}}
\newcommand{\orcidauthor}[2]{{\em {#1}}\hspace*{1mm}\href{https://orcid.org/#2}{\includegraphics[width=1em]{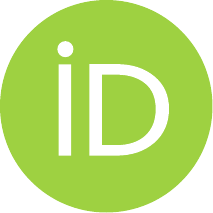}}}

\newif\iftrackchanges
\trackchangesfalse
\newcommand{\add}[1]{\iftrackchanges{\color{OliveGreen}#1}\else#1\fi}
\newcommand{\remove}[1]{\iftrackchanges{\color{BrickRed}[--~#1~--]}\fi}
\newcommand{\changed}[2]{\remove{#1}\iftrackchanges\ \fi\add{#2}}

\makeatletter
\newcommand{\linebreakand}{%
    \end{@IEEEauthorhalign}
    \hfill\mbox{}\par
    \mbox{}\hfill\begin{@IEEEauthorhalign}
}
\makeatother

\newif\ifblind
\blindfalse

\newcommand{\blind}[2]{\ifblind #2\else #1\fi}

\newsavebox{\blindbox}
\newlength{\blindht}
\newlength{\blinddp}
\newcommand{\blindpar}[2]{%
  \ifblind
    \savebox{\blindbox}{\parbox{\linewidth}{#1}}%
    \settoheight{\blindht}{\usebox{\blindbox}}%
    \settodepth{\blinddp}{\usebox{\blindbox}}%
    \addtolength{\blindht}{\blinddp}%
    \begin{minipage}[t][\blindht][c]{\linewidth}\centering #2\end{minipage}%
  \else #1\fi}

\ifblind
  \renewcommand{\orcidauthor}[2]{%
    \phantom{{\em {#1}}\hspace*{1mm}\includegraphics[width=1em]{orcid.pdf}}}
  \renewcommand{\emailaddr}[1]{\phantom{{\small\normalfont\texttt{#1}}}}
\fi

\begin{document}
\title{A Perch out of Water: Few-Shot Detection of Bryde's Whale Vocalisations in False Bay}
% \\\large

\author{
    \IEEEauthorblockN{\orcidauthor{G\"{u}nther Tonitz}{0009-0009-6030-4122}}
    \IEEEauthorblockA{\emailaddr{gtonitz@sun.ac.za}}
    \and
    \IEEEauthorblockN{\orcidauthor{Christiaan M. Geldenhuys}{0000-0003-0691-0235}}
    \IEEEauthorblockA{\emailaddr{cmgeldenhuys@sun.ac.za}}
    \and
    % \IEEEauthorblockN{\orcidauthor{Jaco D.J.J. Versfeld}{0000-0001-8327-2244}}
    % \IEEEauthorblockA{\emailaddr{djjversfeld@sun.ac.za}}
    % \and
    \IEEEauthorblockN{\orcidauthor{Thomas Niesler}{0000-0002-7341-1017}}
    \IEEEauthorblockA{\emailaddr{trn@sun.ac.za}}
    \linebreakand
    \IEEEauthorblockA{
        \blind{\textit{Department of Electrical and Electronic Engineering}}%
              {\texttt{Authors have been removed}}\\
        \blind{\textit{Stellenbosch University}}%
              {\texttt{for the purposes of blind review.}}\\
        \blind{Stellenbosch, South Africa}{\phantom{Stellenbosch, South Africa}}
    }
}

\newcolumntype{P}[1]{>{\centering\arraybackslash}p{#1}}

\maketitle
% \cmg{@GUNTHER, some title suggestions (a) \textit{A Perch out of Water: Few-Shot Baleen Whale Call Detection with Fixed Bioacoustic Embeddings}; (b) \textit{From Tweets to Tones: Transferring Birdsong Embeddings to Baleen Whale Calls}; (c) \textit{Here Comes the Bryde: Few-Shot Detection of Bryde's Whale Vocalisations in False Bay}; (d) \textit{Bryde and Prejudice: Pretrained Bioacoustic Embeddings Meet an Unseen Species}. You could also go isiZulu or isiXhosa with something like \textit{we hear you}?}
% \cmg{@GUNTHER, some subtitle \textit{Few-Shot Detection of Bryde's Whale Vocalisations in False Bay}, \textit{Towards Few-Shot Detection of Baleen Whale Calls, from the Southern Ocean to False Bay}, or \textit{A First Look at Few-Shot Baleen Whale Call Detection in Antarctic and South African Waters}.}

\begin{abstract}
The automated detection of baleen whale calls generally requires large annotated datasets, which are not available for many species and populations.
We investigate whether pretrained bioacoustic embeddings, combined with a prototypical few-shot classifier, allow such calls to be detected and classified from only a handful of labelled examples.
Two embedding networks are compared: Perch, a general-purpose bioacoustic model trained predominantly on birdsong, and WhaleVAD, a task-specific blue- and fin-whale call detector.
% \changed{embedding models}{embedding networks}
Neither model is fine-tuned, and at most ten labelled examples per call type are assumed.
Evaluating on Antarctic blue- and fin-whale recordings, we find that the default Perch input representation is poorly matched to low-frequency whale calls, and that increasing the frame length while restricting the mel filterbank to the frequency range occupied by the calls improves the \acs{aucroc} by 5.8\% absolute when ten examples per call type are available.
% \changed{by 5.4\% absolute}{by 5.8\% absolute with ten examples per call type}
Allowing the classifier to detect each call type independently, so that segments may contain several calls, affords a further 3.3\% improvement and a macro \acs{aucroc} of 0.739, representing a combined gain of 9.1\% absolute.
% A multilabel one-vs-background formulation of the classifier affords a further 3.3\%, for a macro \acs{aucroc} of 0.739 and a combined gain of 9.1\% absolute.
% \changed{affords a further improvement, for a combined gain of 9.1\% absolute}{affords a further 3.3\%, for a macro \acs{aucroc} of 0.739 and a combined gain of 9.1\% absolute}
Both embeddings are applied to Bryde's whale vocalisations recorded in False Bay, South Africa, which neither model has encountered during pretraining.
Here Perch outperforms WhaleVAD for every call type, achieving a macro AUC-ROC of 0.987 against 0.859, despite having a predominantly terrestrial training set.
These results indicate that general-purpose bioacoustic embeddings, paired with an input representation matched to the calls of interest, can support whale call detection for species and populations for which large annotated datasets do not exist.
\end{abstract}

\vspace*{1mm}

%%%%%%%%%%%%%%%%%%%%%%%%%%%%%%%%%%%%%%%%%%%%%%%%%%%%%%
\section{Introduction}
%%%%%%%%%%%%%%%%%%%%%%%%%%%%%%%%%%%%%%%%%%%%%%%%%%%%%%

\Ac{pam} has become a standard tool for the study of marine mammal populations in remote and otherwise inaccessible habitats, owing to its non-invasive nature and relatively modest cost.
In practice, however, \ac{pam} generates large volumes of data in which the signal-to-noise ratio is often low~\autocite{Kowarski2021ReviewPAM}.
% \changed{substantial volumes of data, often suffering from low signal-to-noise ratios}{large volumes of data in which the signal-to-noise ratio is often low}
Manual review of these recordings is therefore very labour-intensive and dependent on specialist knowledge.
This has motivated research into automated detectors for the vocalisations of key species.
\Aclp{dnn} currently dominate this field, and have been shown to match or exceed the performance of an experienced human annotator~\autocite{Miller2023Dcalls}.
% \changed{match or even exceed the performance of experienced human annotators}{match or exceed the performance of an experienced human annotator}
Such systems nevertheless require large annotated datasets for training, which are seldom available in practice~\autocite{kershenbaum2024Guide, Gibb2019Emerging}.

An example for which such data is not available is the Bryde's whale (\textit{Balaenoptera edeni}).
% are one such species~\autocite{constantine2018Future} for which limited data is available.
The South African inshore population, which numbers fewer than \num{1000} individuals, is classified as vulnerable~\autocite{penry2016Conservation}, and remains exposed to vessel strikes and entanglement in fishing gear~\autocite{segre2022Highspeed}.
Effective conservation of this population depends on the ability to monitor its acoustic presence.
To our knowledge, no deep learning detector has been developed for the detection of Bryde's whale calls in False Bay.

\Ac{fsl} offers a means of addressing a scarcity of training examples, by allowing novel classes to be recognised from only a handful of labelled examples~\autocite{vinyals2017MatchingNetworks}.
Metric-based approaches, and prototypical networks in particular~\autocite{snell2017ProtoNet}, have proven especially well suited to \ac{sed}, where performance depends strongly on the quality of the embedding provided to the classifier.
In this paper we investigate whether the embeddings produced by existing pretrained bioacoustic models are appropriate for the few-shot detection and classification of baleen whale calls.
% Metric-based approaches, and prototypical networks in particular~\autocite{snell2017ProtoNet}, have proven especially well suited to \ac{sed}, since they require only a single feed-forward pass at inference.
% Their performance depends mainly on the discriminative quality of the embedding on which they operate.
% In this paper we therefore investigate whether the embeddings produced by existing pretrained bioacoustic models are sufficient to support the few-shot detection and classification of baleen whale calls.

We compare two such models, neither of which is trained or fine-tuned in any way.
\gls{perch}~\autocite{merrienboer2025perch2.0} is a general-purpose bioacoustic model trained predominantly on birdsong, while \gls{wad}~\autocite{geldenhuys2025WhaleVAD} is a task-specific detector trained on blue- and fin-whale calls.
Both are evaluated on the \ac{atbfl} dataset, and on a dataset of Bryde's whale vocalisations representing a species that neither model encountered during training.
Our experiments allow us to draw conclusions regarding the best way to study new types of whale call.
% \changed{for which}{on which}
% \ac{fb}, South Africa, \changed{which neither model has encountered during training}{from a species that neither model encountered during training}

We make the following contributions:
\begin{itemize}
    \item We show that the default \gls{perch} input representation is poorly matched to the low-frequency calls of baleen whales, and propose a matched input representation, which trades time resolution for frequency resolution by increasing the frame length and concentrating the mel filterbank in the band occupied by the calls.
    \item We further introduce a multilabel one-vs-background formulation of the prototypical classifier, which supports segments containing more than one call type.
    \item Finally, we apply deep learning for the first time to the Bryde's whale population of False Bay and find that \gls{perch} transfers considerably better to the unseen Bryde's whale calls than \gls{wad}, despite the latter having been trained on baleen whale vocalisations.
\end{itemize}

% We show that the default \gls{perch} input representation is poorly matched to the low-frequency calls of baleen whales, and propose a matched input representation, which trades time resolution for frequency resolution by increasing the frame length and concentrating the mel filterbank in the band occupied by the calls.
% % We show that the default \gls{perch} input representation is poorly matched to the low-frequency calls of baleen whales, and propose adjustments to the frame length and the mel filterbank which lead to substantial improvements.\cmg{The supervisor asked that the introduction sell the method rather than the numbers, and this sentence ends on the outcome. Suggest: \textit{and propose a matched input representation, which trades time resolution for frequency resolution by lengthening the analysis window and concentrating the mel filterbank in the band occupied by the calls}.}
% We further introduce a multilabel one-vs-background formulation of the prototypical classifier, which supports segments containing more than one call type.
% Finally, we find that \gls{perch} transfers considerably better to the unseen Bryde's whale calls than \gls{wad}, despite the latter having been trained on baleen whale vocalisations.

%%%%%%%%%%%%%%%%%%%%%%%%%%%%%%%%%%%%%%%%%%%%%%%%%%%%%%
\section{Background}
%%%%%%%%%%%%%%%%%%%%%%%%%%%%%%%%%%%%%%%%%%%%%%%%%%%%%%
\vspace{0.5em}

This section summarises few-shot learning and prototypical networks, and describes the architectures of the two pretrained models used as embedding networks in this work.
% \changed{provides a summary of the background regarding}{summarises}
% \changed{that are used}{used}

\vspace{0.25em}
\subsection{Prototypical networks}
\label{ssec:prototypical}
\vspace{0.25em}

\Acf{fsl} refers to the development of models which generalise to novel classes when provided with only a very small number of labelled examples.
% \changed{are able to generalise}{generalise}
Such techniques are useful in applications where the collection or annotation of data is expensive or otherwise difficult and conventional supervised classifier training is therefore not feasible.
Metric-based methods address this by mapping the input to an embedding space in which examples from the same class are clustered together while examples from different classes are well separated.
In this space, a query is classified according to its distance to a small set of labelled references~\autocite{vinyals2017MatchingNetworks}.
Such models are typically trained episodically, so that each training episode mimics the structure of the task encountered at inference.
Since inference requires only a single feed-forward pass, such methods are particularly well suited to detection in long continuous recordings~\autocite{wang2020FSSED}.

\textcite{snell2017ProtoNet} introduced prototypical networks, which assume that an embedding space exists in which each class is well represented by a single \textbf{prototype}, computed as the mean of the small number of available labelled examples, also referred to as the \textbf{support set}.
A query is embedded by the same network and assigned to the class whose prototype lies nearest, using a measure such as the Euclidean distance.
The classifier therefore introduces no trainable parameters of its own, and its performance depends entirely on the discriminative quality of the embedding.
In this work, no training or fine-tuning is performed.
The embedding is obtained from a pretrained model whose weights remain fixed throughout.

\vspace{0.25em}
\subsection{\gls{wad}}
\label{ssec:whalevad}
\vspace{0.25em}

\gls{wad}~\autocite{geldenhuys2025WhaleVAD} is a lightweight \ac{sed} system developed for the per-frame detection of blue- and fin-whale calls, originally inspired by a \ac{vad} framework proposed for speech.
The model operates directly on the \ac{stft}, representing each spectral component by its magnitude together with the sine and cosine of its phase. 
% since the inclusion of phase information was found to improve detection performance.
A linear convolutional layer acts as a learnable filterbank, and is followed by a two-layer \ac{cnn} with max pooling, GELU activation and batch normalisation.
The resulting features are passed through a residually connected bottleneck network, which compresses them to a lower-dimensional representation, and then through a series of depthwise convolutional layers.
These latent features are processed sequentially by a \ac{bilstm} network, after which a linear layer with sigmoid activation produces a posterior probability for each call type at every frame.
The model was trained on the \ac{atbfl} training set using a focal loss in combination with stochastic negative mini-batch undersampling, both of which serve to address the class imbalance of the data.

\vspace{0.25em}
\subsection{\gls{perch}}
\label{ssec:perch}
\vspace{0.25em}

\gls{perch}~\autocite{merrienboer2025perch2.0} is a general-purpose bioacoustic model, trained in a supervised fashion to provide both species classifications and general-purpose embeddings for transfer learning.
% \changed{predictions}{classifications}
The most recent release, Perch 2.0, extends the original bird-only model to a multi-taxa training set of nearly \num{15000} terrestrial and avian species drawn from Xeno-Canto~\autocite{vellinga2015xeno}, iNaturalist~\autocite{inaturalist}, the Animal Sound Archive~\autocite{MfN2016AnimalSoundArchive} and FSD50K~\autocite{fonseca2020fsd50k}.
% \remove{, and is the version used throughout this work, to which we refer simply as \gls{perch}}
This is the version used in all experiments, and is referred to simply as \gls{perch}.
% \add{This is the version used in all experiments, and is referred to simply as \gls{perch}.}
The front-end transforms a \qty{5}{\second} segment of the audio input, sampled at \qty{32}{\kilo\hertz}, into a log-mel spectrogram, which is embedded by an EfficientNet-B3 backbone of approximately \num{12} million parameters.
The backbone produces a spatial embedding with axes corresponding to time, frequency and feature channels.
This is averaged over the two spatial dimensions to obtain a single \num{1536}-dimensional embedding.
Three output layers are attached during training, but only the embedding network is used in this work.
% \changed{heads}{layers}
% \changed{model}{network}

%%%%%%%%%%%%%%%%%%%%%%%%%%%%%%%%%%%%%%%%%%%%%%%%%%%%%%
\section{Literature Review}
%%%%%%%%%%%%%%%%%%%%%%%%%%%%%%%%%%%%%%%%%%%%%%%%%%%%%%
\vspace{0.5em}

This section begins with an overview of previous research into the automated detection and classification of baleen whale calls, with particular attention to the limited work concerning Bryde's whales.
The application of \ac{fsl} to \ac{sed} is then considered, with attention to the comparatively little prior work concerning marine species.
% \changed{together with the comparatively little attention that has been afforded to}{with attention to the comparatively little work concerning}

\vspace{0.25em}
\subsection{Detection and classification of baleen whale calls}
\vspace{0.25em}

Automated detection of baleen whale calls has progressed from matched filtering and spectrogram correlation~\autocite{Mellinger2000SpecCor}, through statistical models such as Gaussian mixture and hidden Markov models, to \aclp{dnn}.
\Acp{cnn} have become the dominant architecture in computational bioacoustics~\autocite{Kowarski2021ReviewPAM}, and \textcite{Miller2023Dcalls} report that a DenseNet-based detector surpassed an experienced human observer in the detection of blue whale D-calls.
Convolutional recurrent architectures, which introduce recurrent layers to capture temporal dependencies, have achieved comparable performance in blue- and fin-whale classification~\autocite{geldenhuys2025WhaleVAD}.
However, such systems depend on large annotated datasets, which are often unavailable in practice~\autocite{kershenbaum2024Guide, Gibb2019Emerging}.

Among baleen whales, the Bryde's whale has received even less attention~\autocite{constantine2018Future}.
% Bryde's whales (\textit{Balaenoptera edeni}) have received far less attention~\autocite{constantine2018Future}.
% The South African inshore population is classified as vulnerable, numbering fewer than \num{1000} individuals~\autocite{penry2016Conservation}, and remains exposed to vessel strikes and entanglement in fishing gear~\autocite{segre2022Highspeed}.\cmg{This sentence duplicates the introduction almost verbatim, with the same two citations.}
Their vocal repertoire has been described for several geographically distinct populations~\autocite{oieson2003Low, heimlich2005Types}, although call types and their rate of occurrence vary substantially between regions.
Isolated populations frequently produce calls that are not described in the existing literature~\autocite{helble2016Swim}.
% \changed{regularly yield}{frequently produce}
% \changed{absent from}{not described in}

Automated detection of the calls of this population has rarely been attempted.
% \add{the calls of }
\textcite{ogundile2020Hidden} and \textcite{babalola2021Detection} both applied hidden Markov models with hand-crafted time-domain features to short pulse calls recorded in False Bay, near Cape Town, South Africa.
\textcite{faustmann2025Brydes} subsequently applied an energy detector to isolate candidate vocalisations, which were then classified by a human annotator, thereby yielding a small annotated dataset.
Elsewhere, \textcite{allen2024Brydes} trained a \ac{cnn} to detect the recently described \textit{Biotwang} call in the North Pacific.
This call type has not, however, been reported in False Bay.
To our knowledge, no deep learning detector has been developed for the vocalisations of the South African inshore Bryde's whale population.

\subsection{Few-shot sound event detection}

% \Ac{fsl} addresses the scarcity of labelled data by training models to recognise novel classes from only a small number of labelled examples, typically by means of an episodic procedure\cmg{Few-shot learning is now defined three times, here and in the introduction and the background, each citing vinyals2017MatchingNetworks, and the single-feed-forward-pass argument likewise appears three times. Repetition} in which each training episode mimics the structure of the inference task~\autocite{vinyals2017MatchingNetworks}.
\textcite{wang2020FSSED} were the first to formalise \ac{fsl} for \ac{sed}, posing it as an open-set detection problem in which a system must localise all occurrences of a target sound within a continuous recording, given only a handful of annotated examples.
The relevance of this formulation to bioacoustics was subsequently recognised by the DCASE challenge, which introduced a dedicated few-shot bioacoustic \ac{sed} task in 2021--2024~\autocite{morfi2021DCASEFSL, Liu2024DCASE}.
% \cmg{just double check the citation}
Metric-based methods, and prototypical networks in particular, have dominated every edition of the challenge~\autocite{nolasco2022DCASEFSL, nolasco2023DCASEFSL}, with the most impactful developments being the shift from segment-level to frame-level prototypes~\autocite{tang2022DCASEFrame} and the use of transductive inference~\autocite{yang2021DCASETransInf}.
% \changed{the challenge throughout}{every edition of the challenge}

The challenge data, however, remains predominantly terrestrial.
A subset of bottlenose dolphin quacks was added in 2022, and pilot whale vocalisations in 2024.
Both proved consistently among the most difficult to process for all submitted systems~\autocite{nolasco2023DCASEFSL, liang2024DCASEDomainGap}.
\textcite{liang2024DCASEDomainGap} attribute this to the mismatch between the underwater and the terrestrial acoustic environment, and report that the \changed{baseline}{reference} network cannot adapt to marine vocalisations without exposure to underwater sound.
Pretrained embeddings offer one means of addressing this.
\textcite{ghani2023GlobBirdEmb} found that classifiers operating on embeddings from models trained on large avian datasets, \gls{perch} among them, consistently outperformed those using general-purpose audio models across a range of taxa, achieving an \ac{aucroc} of \num{0.98} on the \ac{wmmsd} with as few as \num{32} examples per class, against a figure of \num{0.56} achieved by an embedding from VGGish.
% \changed{trained on}{operating on embeddings from models trained on}
% \changed{general-purpose audio models}{those using general-purpose audio models}
\textcite{williams2025ReefSet} extended this investigation to the coral reef domain, but observed that the bird-only \gls{perch} model still outperformed their reef-adapted variant on the same marine mammal recordings.
\textcite{perch2025transfers} further evaluated \gls{perch} on a number of additional marine datasets, including the \ac{noaa} \ac{pipan} dataset, which contains Bryde's whale calls recorded in the North Pacific.
The model performed well on this dataset, but it covers only the \textit{Biotwang} call of that region.
% and contains comparatively few annotated examples of it.
% \changed{While the model performed well on this dataset, it covers}{The model performed well on this dataset, but that dataset covers}
% \changed{comprises}{contains}
% \changed{that class}{it}
These evaluations were nevertheless conducted as closed-set classification over curated audio extracts, rather than as detection within continuous \ac{pam} recordings.
To our knowledge, no few-shot system has been evaluated for the detection of baleen whale calls under the latter conditions.

%%%%%%%%%%%%%%%%%%%%%%%%%%%%%%%%%%%%%%%%%%%%%%%%%%%%%%
\section{Data}
\label{sec:data}
%%%%%%%%%%%%%%%%%%%%%%%%%%%%%%%%%%%%%%%%%%%%%%%%%%%%%%

% Describe both atbfl and false bay datasets
The \acl{atbfl} dataset consists of strongly labelled blue- and fin-whale vocalisations recorded in the Antarctic region of the Southern Ocean.
It was originally released by the \ac{atp} as part of the \ac{iwc-sorp}.
The dataset comprises several site-year subsets, of which only \texttt{casey2017} and \texttt{kerguelen2015} are used in this study.
This decision was motivated by the intention of evaluating the systems across distinct recording environments, since the remaining \ac{atbfl} validation subset was recorded in the same \texttt{kerguelen} site in \num{2014}.
% both by computational constraints and
Both subsets are a part of the validation set, which is fully disjoint from the \ac{atbfl} training set.
This is relevant because \gls{wad} is trained on the ATBFL training set.
% \gt{@trn: Some decisions for WhaleVAD were based on these folds, but similarly we are selecting the perch configuration also on these folds. So the true focus is the unseen performance on False Bay.}
Each subset contains hour-long recordings that have been manually annotated.
The annotations distinguish seven different call types, which are commonly grouped according to their acoustic similarity into the three classes \texttt{bmabz}, \texttt{bp} and \texttt{d}, as proposed by \textcite{schall2024Benchmark}.
The \texttt{bmabz} class comprises the blue whale A-, B- and Z-calls, which co-occur and are narrowband tonal calls lying predominantly between \qtyrange{15}{30}{\hertz}.
The \texttt{bp} class comprises the fin whale \qty{20}{\hertz} pulse, a brief downsweep from \qtyrange{30}{15}{\hertz}, which may be accompanied by an overtone between \qtyrange{80}{120}{\hertz}.
Finally, the \texttt{d} class comprises the blue whale D-call, a downsweep from approximately \qty{120}{\hertz} to \qty{20}{\hertz}, together with the fin whale downsweep between \qtyrange{30}{90}{\hertz}.
% All three classes therefore contain substantial energy below \qty{60}{\hertz}, and both the \texttt{bmabz} class and the fin whale pulse of the \texttt{bp} class lie entirely below the band occupied by the Bryde's whale call types described below.

\Cref{subtbl:recording_info_atbfl} summarises the total recording duration and the number of annotated calls for each subset, while \Cref{subtbl:call_info_atbfl} presents the corresponding per-call duration statistics.
Both subsets exhibit a class imbalance, with annotated whale calls accounting for only \qty{3.49}{\percent} of the total recording duration.
Furthermore, the three call types do not occur with equal frequency, with \texttt{bmabz} calls being the most abundant and also the longest in duration.
% \changed{since annotated whale calls account}{with annotated whale calls accounting}

\begin{figure}
    \centering
    \begin{subfigure}[t]{0.49\columnwidth}
        \centering
        \includegraphics[width=\linewidth]{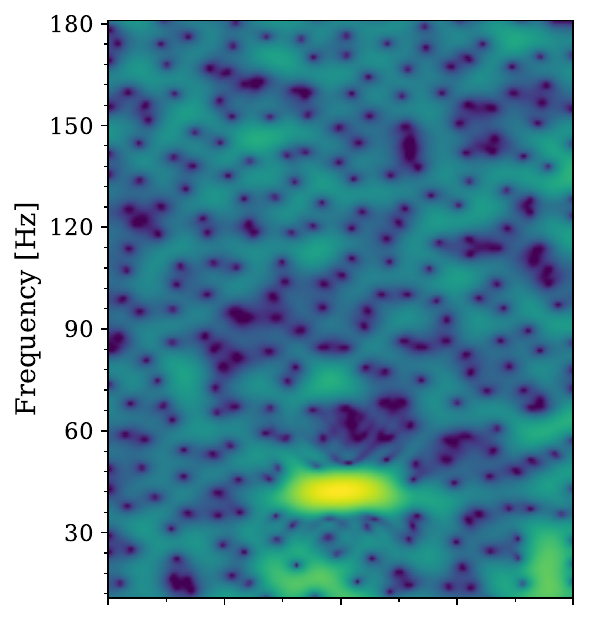}
        \caption{\texttt{ST}}
        \label{fig:call_examples_st}
    \end{subfigure}%
    \hfill%
    \begin{subfigure}[t]{0.42\columnwidth}
        \centering
        \includegraphics[width=\linewidth]{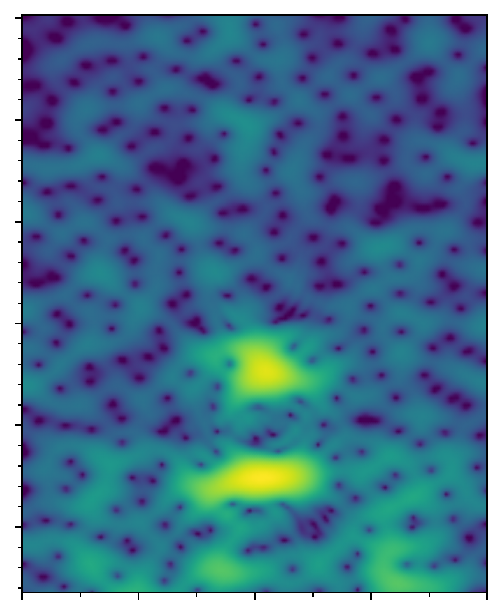}
        \caption{\texttt{MT}}
        \label{fig:call_examples_mt}
    \end{subfigure}

    \vspace{0.5em}

    \hspace{0.05em}
    \begin{subfigure}[t]{0.49\columnwidth}
        \centering
        \includegraphics[width=\linewidth]{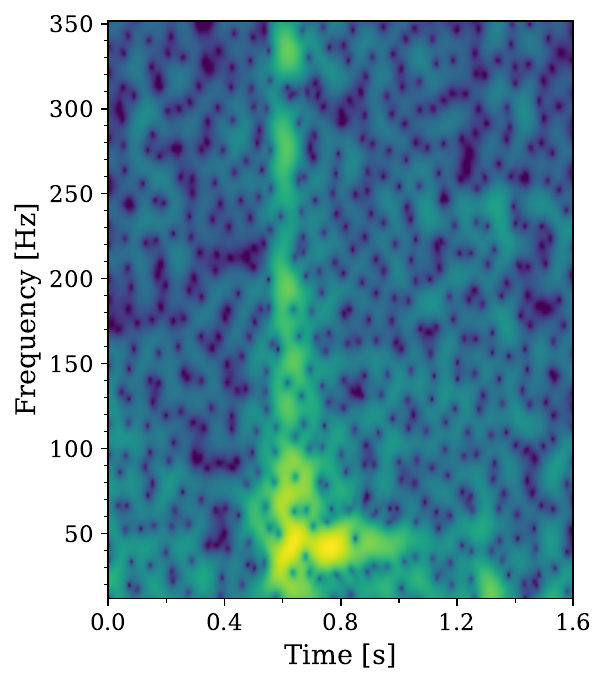}
        \caption{\texttt{BT}}
        \label{fig:call_examples_bt}
    \end{subfigure}%
    \hfill%
    \begin{subfigure}[t]{0.43\columnwidth}
        \centering
        \includegraphics[width=\linewidth]{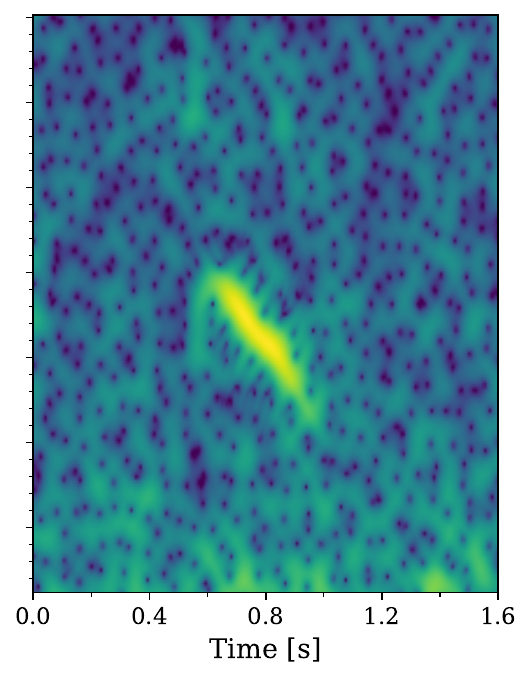}
        \caption{\texttt{TD}}
        \label{fig:call_examples_td}
    \end{subfigure}

    \caption{Spectrogram representations of the four Bryde's whale call types found in the \acl{fb} dataset.}
    \label{fig:call_examples}
    \vspace{-1.25em}
\end{figure}

The \ac{fb} dataset consists of Bryde's whale vocalisations recorded off the coast of South Africa, and was originally published by \textcite{faustmann2025Brydes}.
A continuously-recording hydrophone was deployed in False Bay, near Cape Town, and was retrieved intermittently to recover the recorded data.
% \changed{monitored the environment continuously, with intermittent retrievals to recover the recorded data}{recorded continuously, and was retrieved intermittently to recover the recorded data}
In a subset of these recordings, energy detection was used to isolate segments containing possible vocalisations.
Each of these segments was then assigned to one of four call types by a human labeller.
% \changed{A subset of these recordings was subsequently studied, whereby energy}{In a subset of these recordings, energy}
% \changed{classified as}{assigned to}
The single-tone (\texttt{ST}) call is a short call centred at \qty{42.2}{\hertz}, and is similar to the multi-tone (\texttt{MT}) call, which consists of a comparable low-frequency tone followed by one or more higher tones.
The burst tonal (\texttt{BT}) call consists of a brief burst of wideband noise, with a dominant component usually near \qty{42}{\hertz}.
Finally, the tonal downsweep (\texttt{TD}) call typically begins at approximately \qty{200}{\hertz} and ends near \qty{150}{\hertz}.
\Cref{fig:call_examples} shows an example of each of these call types.
% \add{call }

The \ac{fb} dataset consists of three \qty{74.6}{\hour} recordings, each corresponding to a different period in April \num{2023}.
The dataset originally contained \num{1032} annotated calls, but the use of an energy detector to select segments for annotation means calls may have been missed.
Therefore, the first recording was strongly annotated in its entirety for the purposes of this study, contributing a further \num{1180} calls.
\Cref{subtbl:recording_info_fb} lists the number of calls and the total call duration for each recording.
The strongly annotated recording (\texttt{2023-04-22}) contains Bryde's whale calls for only \qty{0.38}{\percent} of its duration, and therefore exhibits a high class imbalance.
\Cref{subtbl:call_info_fb} summarises the number and the duration of the various call types for each recording.
All four call types are short, with average durations well below \qty{1}{\second}.
In addition, the recordings \texttt{2023-04-26} and \texttt{2023-04-29} contain no \texttt{TD} calls.

%%%%%%%%%%%%%%%%
% Recording info
%%%%%%%%%%%%%%%%
\begin{table}
    \centering
    \caption{
        Summary of the \ac{atbfl} and \ac{fb} recordings used for experimentation, showing the total recording duration (hours), the number of annotated calls, and the total duration of the annotated calls (minutes).%\cmg{@GUNTHER, just double check the math here. First, \texttt{kerguelen2015} is given as 5542 events here, but the per-type counts sum to 2748+1269+1524=5541, so either this figure or the total of 8805 is out by one. \texttt{casey2017} reconciles exactly.}
    }
    \begin{subtable}{\linewidth}
        \centering
        \caption{\acs{atbfl}}
        \begin{tabular}{@{}w{l}{2.0cm}ccc@{}}
            \toprule
            Subset & Duration~(h) & Calls & Total call duration~(min) \\
            \midrule
            \texttt{casey2017}                   & 186.9  & 3263  & 366.7  \\
            \texttt{kerguelen2015}               & 199.9  & 5541  & 443.5  \\
            \midrule
            \textbf{Total}       & \textbf{386.8}  & \textbf{8804} & \textbf{810.2} \\
            \bottomrule
        \end{tabular}
        \label{subtbl:recording_info_atbfl}
    \end{subtable}

    \vspace{1em}

    \begin{subtable}{\linewidth}
        \centering
        \caption{\acs{fb}}
        \begin{tabular}{@{}w{l}{2.0cm}ccc@{}}
            \toprule
            Recording & Duration~(h) & Calls & Total call duration~(min) \\
            \midrule
            \texttt{2023-04-22}                  & 74.6  & 1802  & 17.2  \\
            \texttt{2023-04-26}                  & 74.6  & 342   & 1.4   \\
            \texttt{2023-04-29}                  & 74.6  & 68    & 0.2   \\
            \midrule
            \textbf{Total}       & \textbf{223.7} & \textbf{2212} & \textbf{18.8} \\
            \bottomrule
        \end{tabular}
        \label{subtbl:recording_info_fb}
    \end{subtable}
    \label{tbl:recording_info}
    \vspace{-0.25em}
\end{table}

%%%%%%%%%%%%%%%%%
% Call statistics
%%%%%%%%%%%%%%%%%

\begin{table}
    \centering
    \caption{
        Minimum, maximum and average call duration (seconds), as well as the number of annotations, for each call type present in the \acs{atbfl} and \acs{fb} data presented in \Cref{tbl:recording_info}.
    }
    \begin{subtable}{\linewidth}
        \centering
        \caption{\acs{atbfl}}
        \begin{tabular}{@{}w{l}{2.0cm}w{c}{1.0cm}|ccc|c@{}}
            \toprule
            \multirow{2}{*}{Subset} &\multirow{2}{*}{Type} & \multicolumn{3}{c|}{Duration (s)} & \multirow{2}{*}{Count} \\
                        & & Min  & Max  & Avg.  & \\
            \midrule
            \texttt{casey2017}      & \texttt{bmabz}    & 1.36 & 36.62 & 8.29 & 2418 \\
            \texttt{casey2017}      & \texttt{bp}    & 0.94 & 2.09 & 1.37 & 292 \\
            \texttt{casey2017}      & \texttt{d}    & 1.10 & 7.01 & 2.80 & 553 \\
            {} & {} & {} & {} & {} & {} \\
            \texttt{kerguelen2015}  & \texttt{bmabz}    & 3.32 & 20.39 & 7.56 & 2748 \\
            \texttt{kerguelen2015}  & \texttt{bp}    & 0.61 & 2.83 & 1.58 & 1269 \\
            \texttt{kerguelen2015}  & \texttt{d}    & 0.37 & 6.88 & 2.52 & 1524 \\
            \bottomrule
        \end{tabular}
        \label{subtbl:call_info_atbfl}
    \end{subtable}

    \vspace{1em}

    \begin{subtable}{\linewidth}
        \centering
        \caption{\acs{fb}}
        \begin{tabular}{@{}w{l}{2.0cm}w{c}{1.0cm}|ccc|c@{}}
            \toprule
            \multirow{2}{*}{Recording} &\multirow{2}{*}{Type} & \multicolumn{3}{c|}{Duration (s)} & \multirow{2}{*}{Count} \\
                        & & Min  & Max  & Avg.  & \\
            \midrule
            \texttt{2023-04-22}  & \texttt{BT}    & 0.42 & 2.58 & 0.69 & 171 \\
            \texttt{2023-04-22}  & \texttt{MT}    & 0.27 & 1.22 & 0.57 & 452 \\
            \texttt{2023-04-22}  & \texttt{ST}    & 0.23 & 1.35 & 0.54 & 1015 \\
            \texttt{2023-04-22}  & \texttt{TD}    & 0.45 & 0.95 & 0.67 & 164 \\
            {} & {} & {} & {} & {} & {} \\
            \texttt{2023-04-26}  & \texttt{BT}    & 0.35 & 0.55 & 0.45 & 9 \\
            \texttt{2023-04-26}  & \texttt{MT}    & 0.13 & 0.52 & 0.27 & 91 \\
            \texttt{2023-04-26}  & \texttt{ST}    & 0.10 & 0.65 & 0.23 & 242 \\
            \texttt{2023-04-26}  & \texttt{TD}    & -    & -    & -    & 0 \\
            {} & {} & {} & {} & {} & {} \\
            \texttt{2023-04-29}  & \texttt{BT}    & 0.10 & 0.51 & 0.26 & 11 \\
            \texttt{2023-04-29}  & \texttt{MT}    & 0.10 & 0.32 & 0.18 & 25 \\
            \texttt{2023-04-29}  & \texttt{ST}    & 0.10 & 0.26 & 0.15 & 32 \\
            \texttt{2023-04-29}  & \texttt{TD}    & -    & -    & -    & 0  \\
            \bottomrule
        \end{tabular}
        \label{subtbl:call_info_fb}
    \end{subtable}

    \label{tbl:call_info}
    \vspace{-1em}
\end{table}

%%%%%%%%%%%%%
% Call figure
%%%%%%%%%%%%%

%%%%%%%%%%%%%%%%%%%%%%%%%%%%%%%%%%%%%%%%%%%%%%%%%%%%%%
\section{Experimental Structure}
%%%%%%%%%%%%%%%%%%%%%%%%%%%%%%%%%%%%%%%%%%%%%%%%%%%%%%

In the following, we first describe the segmentation and labelling procedure applied to both datasets.
We then present the two pretrained models that are used as embedding networks, followed by the few-shot classifier that operates on the resulting embeddings.
Finally, we describe the adjustments made to the \gls{perch} input representation, as well as the evaluation protocol used throughout.

\vspace{-0.25em}
\subsection{Segmentation and labelling}

All experiments consider \qty{5}{\second} fixed-length segments for classification.
This interval was chosen to match the input duration expected by \gls{perch}.
A hop of \qty{4}{\second} between successive segments ensures an inter-segment overlap.
A segment is labelled with a particular call type when its \textit{purity}, defined as the fraction of the segment associated with an annotated call of that type, exceeds \num{0.02}, corresponding to \qty{100}{\milli\second} of call activity.
Segments that do not satisfy this criterion for any call type are labelled as \texttt{negative}.
Since more than one call may occur within a single segment, a segment may carry several labels, and the classification task is therefore both multiclass and multilabel.
%\cmg{@GUNTHER, we never report how often this actually happens. The multilabel formulation is claimed as a contribution, so one number here, the percentage of positive segments carrying more than one label, is the cheapest possible defence of it. Do you have that figure?}
% \changed{inherit}{carry}
Of the positive segments in the \ac{atbfl} and \ac{fb} datasets, \qty{4.15}{\percent} and \qty{14.66}{\percent} are multilabel, respectively. 
For the \ac{fb} dataset, the \texttt{TD} call is excluded from all experiments, since it occurs in only one of the three recordings and independent support samples cannot therefore be drawn from the remaining recordings.
% \changed{can therefore not}{cannot therefore}

\subsection{Models}

Two pretrained models are compared in this work.
In neither case are the model weights trained or fine-tuned.

\gls{perch}, as discussed in \Cref{ssec:perch}, is a general-purpose bioacoustic model trained predominantly on birdsong.
The model processes a \qty{5}{\second} audio input, sampled at \qty{32}{\kilo\hertz}, to produce a \num{1536}-dimensional embedding.
Both datasets are therefore upsampled to \qty{32}{\kilo\hertz}, to match the expected input rate while preserving the original frequency content.
% \remove{in order }

\gls{wad}, as discussed in \Cref{ssec:whalevad}, is a frame-level blue- and fin-whale call detector trained on the \ac{atbfl} training set.
For each frame, computed every \qty{20}{\milli\second}, the model computes a posterior probability for each of the seven original \ac{atbfl} call types.
When evaluated on the \ac{atbfl} dataset, \gls{wad} is therefore a fully supervised detector and can serve as a topline.
Its per-frame posterior probabilities are grouped into the three collapsed classes (see \Cref{sec:data}) by taking the maximum probability within each group.
Subsequently, a threshold may be applied to obtain per-frame binary detections.
A segment is labelled with a given call type when these detections satisfy the same minimum purity criterion of \num{0.02} used for \gls{perch}.
% \cmg{@GUNTHER, the threshold applied to the per-frame posteriors is never given. What value was used, and how was it chosen? Or is this the fancy thresholds. Not that it matters all that much.}
Since \gls{wad} has not been exposed to Bryde's whale vocalisations, it cannot be applied as a detector to the \ac{fb} dataset. %\cmg{@GUNTHER, this is the objection most likely to sink the paper and it is answerable in one clause. Nowhere do we state that the queried \texttt{casey2017} and \texttt{kerguelen2015} recordings are disjoint from the \acs{atbfl} partition \gls{wad} was trained on. If they are not, the topline is measured on seen material and the reported gaps understate it. Were they held out? I remember we talked about this.}
Instead, we use the model as an embedding network by mean-pooling the \num{256}-dimensional output of its recurrent layer over all frames within a segment.
Informal experiments found mean-pooling to outperform max-pooling.
These embeddings are subsequently used in exactly the same way as those produced by \gls{perch}.

\subsection{Few-shot classifier}

Classification is performed by a prototypical classifier, as discussed in \Cref{ssec:prototypical}, operating on the embeddings produced by the two pretrained models.
% \changed{network}{classifier}
% \changed{which uses the embeddings produced by the pretrained models as an embedding network}{operating on the embeddings produced by the two pretrained models}
For each of the $C$ call types, a support set of $k$ segments is drawn, together with a further $k$ segments from the negative class.
A prototype is formed for each class by obtaining the mean of these support embeddings, yielding $C$ class prototypes $\mathbf{p}_c$ along with a negative prototype $\mathbf{p}_0$.
A query segment with embedding $\mathbf{z}$ then yields $C+1$ Euclidean distances $d_c = \lVert \mathbf{z} - \mathbf{p}_c \rVert_2$.
In the \textit{multiclass} configuration, the softmax of the negated distances is taken to obtain a probability vector over the $C+1$ classes, and the class with the highest probability is the classification.
% \changed{negative}{negated}
% \changed{prediction}{classification}

Because a segment may carry more than one label, we further consider a \textit{multilabel} configuration, in which the distances are converted into per-class scores using a \textit{one-vs-background} formulation,
\vspace{-0.25em}
\begin{equation}
    s_c = d_0 - d_c ,
    \label{eq:one_vs_background}
\end{equation}
\vspace{-0.25em}
whose sign indicates whether the query lies closer to the class prototype ($s_c \ge 0$) or to the negative prototype ($s_c < 0$).
% \changed{which measures whether}{whose sign indicates whether}
This allows multiple labels to be assigned to a single segment.

\subsection{\gls{perch} input representation}

The default front-end configuration of \gls{perch} is a poor fit for baleen whale calls.
It computes a log-magnitude spectrogram from a \qty{20}{\milli\second} window (\num{640} samples at \qty{32}{\kilo\hertz}) with a \qty{10}{\milli\second} hop, which is projected onto \num{128} mel bands spanning from \qtyrange{60}{16000}{\hertz}.
Since the lower edge of the first mel filter lies above the fundamental frequency of most of the call types (\Cref{sec:data}), this energy is discarded entirely.
%\cmg{@GUNTHER, this cross-reference is wrong: \texttt{tbl:call\_info} reports durations and counts, no frequencies at all. More importantly, the \acs{atbfl} call frequencies are never given anywhere in the paper, so the claim that the calls lie below \qty{60}{\hertz} is unsupported for \texttt{bmabz}, \texttt{bp} and \texttt{d}. Can we add the frequency range of the three call types?}
Furthermore, a \qty{20}{\milli\second} window offers poor frequency resolution, with a bin spacing of \qty{50}{\hertz}, in the lower bands, where the different call types must be distinguished.
% \changed{very poor}{poor}
% \add{, with a bin spacing of \qty{50}{\hertz},}

The first experiment therefore adjusts the front-end configuration of \gls{perch} by varying the parameters of the mel spectrogram, while the number of mel bands is kept at \num{128} so that the subsequent convolutional layers receive an input of unchanged dimensionality.
This subsequently remains fixed for all further experiments.
First, the frame length is increased from the default of \num{640} samples to a value between \num{1024} and \num{32768}.
In addition, the mel filterbank is extended to span from \qtyrange{0}{16000}{\hertz}.
% \changed{from \num{640} to \num{32768} samples, increasing in powers of two}{from the default of \num{640} samples, and then in powers of two from \num{1024} to \num{32768}}
The upper frequency limit of the filterbank is then lowered from \qty{16000}{\hertz} to \qty{1000}{\hertz}, thereby concentrating the available mel bands in the frequency range occupied by the whale calls and increasing the effective frequency resolution.
% \changed{further restricted to \qtyrange{0}{1000}{\hertz}}{lowered from \qty{16000}{\hertz} to \qty{1000}{\hertz}}

\vspace{-0.4em}
\subsection{Performance evaluation}

Experiments are evaluated under a leave-one-recording-out scheme.
Every segment is used as a query in turn, and its support set of $k$ segments per class is sampled without replacement from recordings other than the one from which the query originates.
This ensures that a query and its supports never share a recording, and therefore that the classifier cannot exploit recording-specific characteristics.
Each query is evaluated against \num{100} independent support draws, and the support set size is varied over $k = 1, \dots, 10$.
% \changed{swept}{varied}
The two \ac{atbfl} subsets are evaluated independently and then averaged to obtain an overall score for the dataset.
For the \ac{fb} dataset, only the strongly annotated \texttt{2023-04-22} recording is queried, since it is the only recording whose unannotated segments can be relied upon to be true negatives.

Performance is reported in terms of the per-class \ac{aucroc} and \ac{ap}, computed against the multilabel targets over all query segments.
Overall scores across classes are computed as the macro average.
Finally, scores are averaged over the \num{100} repeated draws to obtain the mean and standard deviation, which are reported for all experiments.
% \changed{which are used for final evaluation}{reported for all experiments}

\vspace{-0.5em}
%%%%%%%%%%%%%%%%%%%%%%%%%%%%%%%%%%%%%%%%%%%%%%%%%%%%%%
\section{Results}
%%%%%%%%%%%%%%%%%%%%%%%%%%%%%%%%%%%%%%%%%%%%%%%%%%%%%%
\vspace{-0.5em}

In the following, we first present the effect of the adjustments made to the \gls{perch} input representation, because these are applied in all subsequent experiments.
% \changed{subsequently applied throughout}{applied in all subsequent experiments}
We then compare the resulting few-shot systems on the \ac{atbfl} dataset, where the fully supervised \gls{wad} model serves as a topline.
Finally, we evaluate both pretrained models as embedding networks using the \ac{fb} dataset, on which neither has been trained.
% \changed{for which neither has seen any training data}{on which neither has been trained}

\subsection{\gls{perch} input representation}

\Cref{tbl:perch_adjustment_results} presents the effect of the adjustments made to the \gls{perch} front-end, evaluated on the \ac{atbfl} dataset in the multiclass configuration with $k = \num{5}$.
Lowering the lower edge of the mel filterbank from \qty{60}{\hertz} to \qty{0}{\hertz}, while retaining the default frame length, improves the \ac{aucroc} by \qty{1.1}{\percent} absolute and leaves the \ac{ap} unchanged.
% \changed{Extending}{Lowering}
Although the default configuration discards the fundamental frequency of most call types, the recovered energy is still spread over too few mel bands to be of substantial benefit to the model.

Once the filterbank is restricted to lie between \qty{0}{\hertz} and \qty{1000}{\hertz}, performance improves consistently with increasing frame length, peaking at \num{8192} samples (\qty{256}{\milli\second} at \qty{32}{\kilo\hertz}) with an \ac{aucroc} of \num{0.690} and an \ac{ap} of \num{0.162}.
%  \qtyrange[range-phrase = \text{ and }]{0}{1000}{\hertz}
This corresponds to absolute improvements of \qty{5.4}{\percent} and \qty{1.6}{\percent} over the default configuration.
No further gains are observed beyond \num{8192} samples.
This configuration is used for all subsequent experiments, and is referred to as \textit{Perch + adjustment}.
It was selected on the \acs{atbfl} dataset and is applied without further modification to the \acs{fb} dataset.
% so the \acs{fb} results are not subject to this selection.
% \cmg{A reviewer will object that the front-end was selected on the same \acs{atbfl} data on which the gain is then reported, over fifteen candidates, and in the multiclass configuration while the deployed system is multilabel. The honest defence is strong and costs one sentence: \textit{It was selected on the \acs{atbfl} dataset and applied without further modification to the \acs{fb} dataset, so the \acs{fb} results are not subject to this selection.}}

%%%%%%%%%%%%%%%%%%%
% Perch Adjustments
%%%%%%%%%%%%%%%%%%%
\begin{table}
    \centering
    \caption{
        Effect of the modification of the \gls{perch} front-end configuration on classification performance measured on the \acs{atbfl} dataset, using the multiclass configuration with $k = \num{5}$.
        The frame length, the frequency range spanned by the mel filterbank, and the resulting macro-averaged \acs{aucroc} and \acs{ap} are shown.
    }
    \begin{tabular}{@{}lcccc@{}}
        \toprule
        \multirow{2}{*}{Frame length} & \multicolumn{2}{c}{Frequency (\si{\hertz})} & \multirow{2}{*}{\acs{aucroc}} & \multirow{2}{*}{\acs{ap}} \\
        {} & Min & Max & {} & {} \\
        \midrule
        640\textsuperscript{\dag}   & 60    & 16000 & 0.636             & 0.146 \\
        \midrule
        640                         & 0     & 16000 & 0.647             & 0.146 \\
        1024                        & 0     & 16000 & 0.620             & 0.116 \\
        2048                        & 0     & 16000 & 0.534             & 0.035 \\
        4096                        & 0     & 16000 & 0.538             & 0.035 \\
        8192                        & 0     & 16000 & 0.558             & 0.064 \\
        16384                       & 0     & 16000 & 0.564             & 0.064 \\
        32768                       & 0     & 16000 & 0.569             & 0.065 \\
        \midrule
        640                         & 0     & 1000  & 0.618             & 0.082 \\
        1024                        & 0     & 1000  & 0.629             & 0.088 \\
        2048                        & 0     & 1000  & 0.664             & 0.106 \\
        4096                        & 0     & 1000  & 0.683             & 0.149 \\
        8192                        & 0     & 1000  & \textbf{0.690}    & \textbf{0.162} \\
        16384                       & 0     & 1000  & 0.682             & 0.161 \\
        32768                       & 0     & 1000  & 0.688             & 0.150 \\
        \bottomrule
        \multicolumn{5}{@{}l@{}}{\footnotesize \textsuperscript{\dag} Default input representation of Perch.}
    \end{tabular}
    \label{tbl:perch_adjustment_results}
    \vspace{-2.0em}
\end{table}

\vspace{-0.5em}
\subsection{\acs{atbfl} dataset}

\Cref{tbl:final_results_atbfl} presents the per-class and macro-averaged results obtained on the \ac{atbfl} dataset for $k = \num{1}$, \num{5} and \num{10}.
The adjusted input representation improves the macro \ac{aucroc} at every support set size, with absolute gains of \qty[drop-zero-decimal=false]{4.0}{\percent}, \qty{5.4}{\percent} and \qty{5.8}{\percent} respectively.
Almost all of this gain arises from the \texttt{bp} call, whose \ac{aucroc} rises from \num{0.783} to \num{0.950} at $k = \num{10}$, an absolute improvement of \qty{16.7}{\percent}.
% \changed{Most all of the improvement is attributed to}{Almost all of this gain arises from}
The \ac{aucroc} of the \texttt{d} call also improves, by \qty{3.9}{\percent} absolute, while that of the \texttt{bmabz} call falls by \qty{3.4}{\percent} absolute.
% \changed{The \texttt{d} call also improves by \qty{3.9}{\percent}, while the \texttt{bmabz} call deminishes by \qty{3.4}{\percent}}{The \ac{aucroc} of the \texttt{d} call also improves, by \qty{3.9}{\percent} absolute, while that of the \texttt{bmabz} call falls by \qty{3.4}{\percent} absolute}.

The multilabel configuration recovers the \texttt{bmabz} call to an \ac{aucroc} of \num{0.607}, above the \num{0.579} obtained with the unmodified system, and achieves the highest macro \ac{aucroc} at every value of $k$, reaching \num{0.739} at $k = \num{10}$.
% \changed{further improves the performance on the \texttt{bmabz} call type}{recovers the \texttt{bmabz} call to an \ac{aucroc} of \num{0.607}, above the \num{0.579} obtained with the unmodified system,}
The improvement over the multiclass configuration is concentrated in the \texttt{bmabz} and \texttt{d} calls, rising by \qty{6.2}{\percent} and \qty{4.4}{\percent} absolute respectively.
% \changed{improving}{rising}
% \add{absolute }
The multiclass configuration nevertheless retains a small improvement in \ac{ap}, achieving \num{0.176} against \num{0.174} at $k = \num{10}$.
The multilabel deficit is confined entirely to the \texttt{bp} call, whose \ac{ap} falls from \num{0.350} to \num{0.295}, while both \texttt{bmabz} and \texttt{d} improve in \ac{ap} as well as in \ac{aucroc}.
% \cmg{Interesting, the multilabel deficit is confined entirely to \texttt{bp}, whose \acs{ap} falls from 0.350 to 0.295, while both \texttt{bmabz} and \texttt{d} improve in \acs{ap} as well as in \acs{aucroc}. That single loss outweighs the two gains in the macro average and explains the whole divergence. Worth a sentence?}
Taken together, the adjusted input representation and the multilabel formulation yield a \qty{9.1}{\percent} absolute improvement in macro \ac{aucroc} over the unmodified \gls{perch} system at $k = \num{10}$, a relative improvement of \qty[drop-zero-decimal=false]{14.0}{\percent}.

\begin{figure}
    \centering
    \includegraphics[width=\columnwidth]{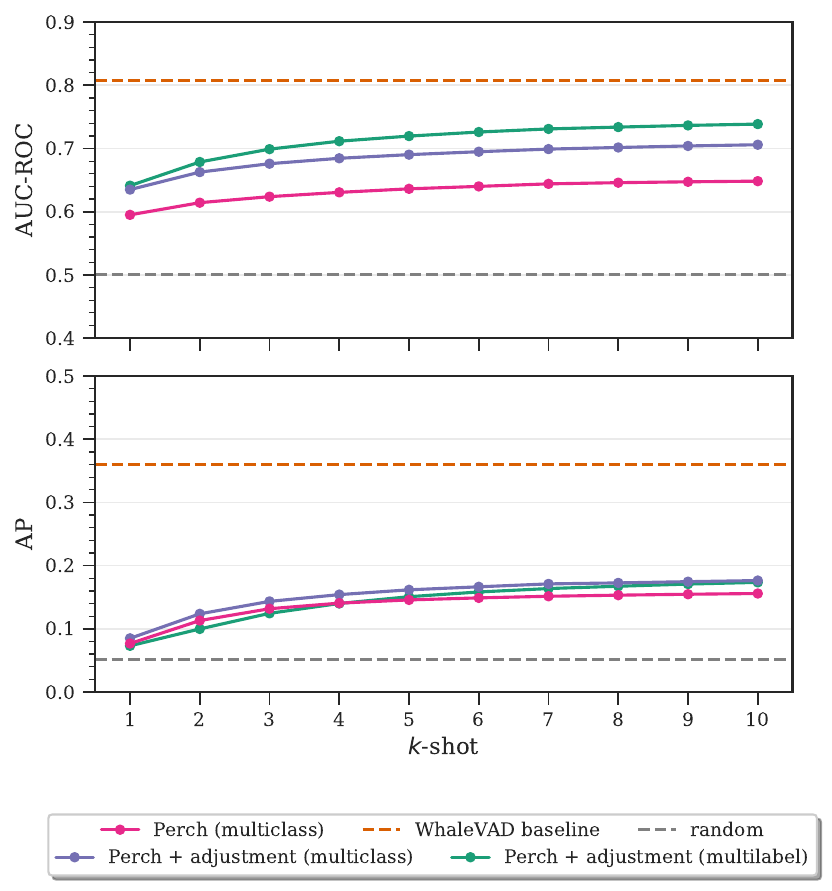}
    \caption{
        Macro-averaged \acs{aucroc} and \acs{ap} on the \acs{atbfl} dataset as a function of the support set size $k$.
        %\cmg{@GUNTHER, the plot axes are labelled \textit{ROC-AUC} and \textit{mAP} while the prose, tables and glossary use \textit{AUC-ROC} and \textit{AP}, and the \acs{fb} figure labels the same axis \textit{Average precision}. Could the figures be regenerated with the labels matched to the text? }
        The dashed lines show the performance of the fully supervised \gls{wad} topline (orange) and a random classifier (grey).
    }
    \label{fig:atbfl_ksweep}
    \vspace{-1.0em}
\end{figure}

\Cref{fig:atbfl_ksweep} shows how performance varies as the support set size is increased from $k = \num{1}$ to $k = \num{10}$.
All three configurations improve with $k$ in terms of both metrics over the range considered.
% All three systems improve monotonically with $k$.
% \cmg{Monotonicity can only be checked from the tables at $k=1,5,10$, where it does hold. Strict monotonicity at every step, and in which metric, is a claim only the figure can support, and the front-end table shows this family of curves is not always monotone. Suggest \textit{All three configurations improve with $k$ in both metrics over the range considered}, unless the figure really is strictly monotone throughout.}
Of the \qty{9.8}{\percent} total absolute gain in \ac{aucroc} achieved by the multilabel configuration, \qty{7.9}{\percent} is obtained between $k = \num{1}$ and $k = \num{5}$.
% \changed{The increase from $k = \num{1}$ to $k = \num{5}$ accounts for \qty{7.9}{\percent} of the \qty{9.8}{\percent} total absolute gain in \ac{aucroc} achieved by the multilabel system}{Of the \qty{9.8}{\percent} total absolute gain in \ac{aucroc} achieved by the multilabel configuration, \qty{7.9}{\percent} is obtained between $k = \num{1}$ and $k = \num{5}$}.
The multilabel configuration attains the highest macro AUC-ROC at every value of $k$ considered, and the adjusted multiclass configuration the highest AP, although the multilabel AP falls below that of the unmodified \gls{perch} system at $k=1$.
% The relative ordering of the three systems also remains stable across all values of $k$, with the multilabel system achieving the highest \ac{aucroc} and the adjusted multiclass system achieving slightly higher \ac{ap} throughout.\cmg{This is contradicted by the table for \acs{ap}. At $k=1$ the ordering is adjusted multiclass 0.085, unmodified 0.077, multilabel 0.073; at $k=5$ and $k=10$ the multilabel configuration moves ahead of the unmodified one. The ordering is therefore not stable. Suggest \textit{The multilabel configuration attains the highest macro AUC-ROC at every value of $k$ considered, and the adjusted multiclass configuration the highest AP, although the multilabel AP falls below that of the unmodified system at $k=1$}. Or have I missed something?}

For the \ac{atbfl} data, we see that none of the few-shot systems approach the fully supervised \gls{wad} topline, which achieves an \ac{aucroc} of \num{0.808} and an \ac{ap} of \num{0.36}.
At $k = \num{10}$, a difference of \qty{7}{\percent} in \ac{aucroc} and \qty{19}{\percent} in \ac{ap} remains.
The absolute \ac{ap} values are low throughout, but must be interpreted with the class imbalance of the dataset in mind. 
The expected \ac{ap} of an uninformative ranking is equal to the positive rate of the class, whereas the expected \ac{aucroc} is always \num{0.5}.
Annotated calls occupy only \qty{3.49}{\percent} of the \ac{atbfl} recordings, and the \ac{ap} of \num{0.174} achieved by the best few-shot system therefore remains well above chance, as does the \num{0.36} achieved by the topline.
The two metrics consequently convey different information, and \ac{ap} in particular cannot be compared across classes or datasets whose positive rates differ.
% The consistently low absolute \ac{ap} values are a further reflection of the severe class imbalance of the dataset.\cmg{The supervisor asked that this relationship be stated rather than assumed. The precise version is short: the expected \acs{ap} of an uninformative ranking equals the positive rate of the class, whereas the expected \acs{aucroc} is 0.5 irrespective of it. That one fact explains the low \acs{atbfl} \acs{ap} values, the 0.174 against a topline of 0.36, and why the pooled \acs{fb} detection \acs{ap} is so much higher than the macro per-class figure.}

\begin{figure}
    \centering
    \includegraphics[width=\columnwidth]{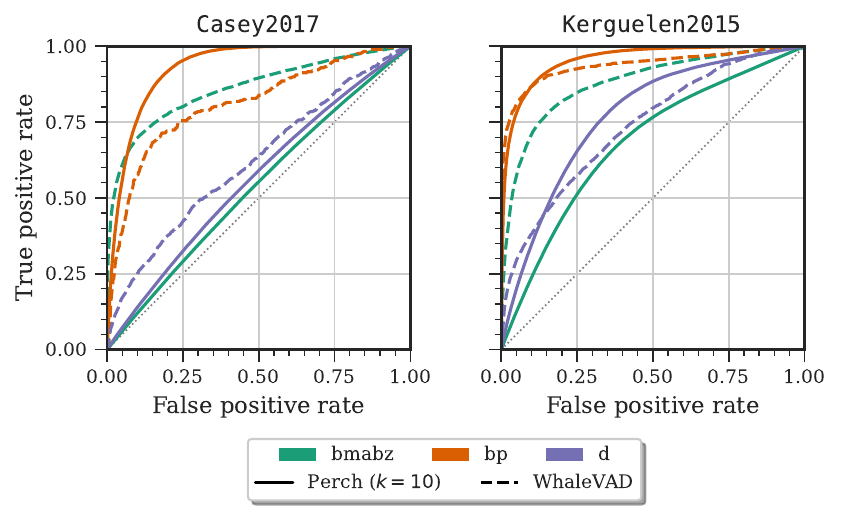}
    \caption{
        Per-class \acs{roc} curves for the \texttt{casey2017} and \texttt{kerguelen2015} subsets, comparing the best few-shot system at $k = \num{10}$ with the fully supervised \gls{wad} model.
    }
    \label{fig:atbfl_roc}
    \vspace{-1.5em}
\end{figure}

\Cref{fig:atbfl_roc} shows the per-class \acs{roc} curves obtained for the two \ac{atbfl} subsets, comparing the best few-shot system at $k = \num{10}$ with \gls{wad}.
\gls{wad} retains a clear advantage for the \texttt{bmabz} and \texttt{d} calls on both subsets.
For the \texttt{bp} call, however, the few-shot classifier is superior.
% , which is consistent with the per-class scores reported in \Cref{tbl:final_results_atbfl}.\cmg{This cross-reference cannot support the claim: \texttt{tbl:final\_results\_atbfl} contains no \gls{wad} rows at all, so it cannot corroborate a comparison against \gls{wad}. The only \gls{wad} \acs{atbfl} numbers in the paper are the two macro values in the preceding paragraph. Either add a \gls{wad} topline row to the table, which would also give the 0.808 and 0.36 a proper home, or drop the clause.}
% \changed{system}{classifier}

While the fully supervised WhaleVAD topline remains ahead, the margin is modest given that the prototypical classifier observes at most ten labelled examples per call type, while WhaleVAD was trained on the full ATBFL training set containing at least \num{1646} examples per call type and over \num{58000} examples in total.
% Whilst the \gls{wad} model remains dominant, the performance of the few-shot classifier remains notable given that the \gls{perch} model has never been trained on marine data, and only uses 10 or fewer samples when \gls{wad} has been trained on substantial blue- and fin-whale calls.\cmg{Suggest: \textit{The fully supervised WhaleVAD topline remains ahead, but the margin is modest given that the prototypical classifier observes at most ten labelled examples per call type, while WhaleVAD was trained on the full ATBFL training set.} }

%%%%%%%%%%%%%
% ATBFL table
%%%%%%%%%%%%%
\begin{table}
    \centering
    \caption{
        Per-class and macro-averaged (\texttt{all}) \acs{aucroc} and \acs{ap} on the \acs{atbfl} dataset, for support set sizes $k = \num{1}$, \num{5} and \num{10}.
        Scores are reported as the mean and standard deviation over \num{100} independent support draws.
        For the \gls{wad} topline, there are no repeated draws and therefore no standard deviations.
    }
    \begin{tabular}{@{}ccl|cc|cc@{}}
        \toprule
        {} & \multirow{2}{*}{Model} & \multirow{2}{*}{Type} & \multicolumn{2}{c|}{\acs{aucroc}} & \multicolumn{2}{c}{\acs{ap}} \\
        {} & {} & {} & Avg. & Std. & Avg. & Std. \\
        \midrule
        \multirow{14}{*}{\rotatebox[origin=c]{90}{}}
            & \multirow{4}{*}{\shortstack{\gls{wad}\\(topline)}}
                              & \texttt{bmabz}    & 0.871           & -- & 0.636             & -- \\
                            & & \texttt{bp}       & 0.874           & -- & 0.356             & -- \\
                            & & \texttt{d}        & 0.678           & -- & 0.088             & -- \\
                            & & \texttt{all}      & 0.808           & -- & 0.360             & -- \\
        \midrule
        \multirow{14}{*}{\rotatebox[origin=c]{90}{$k = 1$}}
            & \multirow{4}{*}{\shortstack{Perch\\(multiclass)}}
                              & \texttt{bmabz}    & 0.527           & 0.003 & 0.126             & 0.001 \\
                            & & \texttt{bp}       & 0.727           & 0.006 & 0.076             & 0.003 \\
                            & & \texttt{d}        & 0.531           & 0.005 & 0.029             & 0.001 \\
                            & & \texttt{all}      & 0.595           & 0.003 & 0.077             & 0.001 \\
            & {} & {} & {} & {} & {} & {} \\
            & \multirow{4}{*}{\shortstack{Perch + adjustment\\(multiclass)}}
                              & \texttt{bmabz}    & 0.511           & 0.003 & 0.120             & 0.001 \\
                            & & \texttt{bp}       & \textbf{0.851}  & 0.004 & \textbf{0.105}    & 0.004 \\
                            & & \texttt{d}        & 0.543           & 0.005 & 0.030             & 0.001 \\
                            & & \texttt{all}      & 0.635           & 0.002 & \textbf{0.085}    & 0.002 \\
            & {} & {} & {} & {} & {} & {} \\
            & \multirow{4}{*}{\shortstack{Perch + adjustment\\(multilabel)}}
                              & \texttt{bmabz}    & \textbf{0.538}  & 0.003 & \textbf{0.130}    & 0.001 \\
                            & & \texttt{bp}       & 0.813           & 0.005 & 0.055             & 0.001 \\
                            & & \texttt{d}        & \textbf{0.572}  & 0.005 & \textbf{0.034}    & 0.001 \\
                            & & \texttt{all}      & \textbf{0.641}  & 0.003 & 0.073             & 0.001 \\
        \midrule
        \multirow{14}{*}{\rotatebox[origin=c]{90}{$k = 5$}}
            & \multirow{4}{*}{\shortstack{Perch\\(multiclass)}}
                              & \texttt{bmabz}    & 0.558           & 0.002 & 0.137             & 0.001 \\
                            & & \texttt{bp}       & 0.786           & 0.005 & 0.268             & 0.004 \\
                            & & \texttt{d}        & 0.564           & 0.005 & 0.033             & 0.001 \\
                            & & \texttt{all}      & 0.636           & 0.002 & 0.146             & 0.001 \\
            & {} & {} & {} & {} & {} & {} \\
            & \multirow{4}{*}{\shortstack{Perch + adjustment\\(multiclass)}}
                              & \texttt{bmabz}    & 0.531           & 0.002 & 0.126             & 0.001 \\
                            & & \texttt{bp}       & \textbf{0.942}  & 0.002 & \textbf{0.321}    & 0.006 \\
                            & & \texttt{d}        & 0.598           & 0.005 & 0.039             & 0.001 \\
                            & & \texttt{all}      & 0.690           & 0.002 & \textbf{0.162}    & 0.002 \\
            & {} & {} & {} & {} & {} & {} \\
            & \multirow{4}{*}{\shortstack{Perch + adjustment\\(multilabel)}}
                              & \texttt{bmabz}    & \textbf{0.588}  & 0.002 & \textbf{0.154}    & 0.001 \\
                            & & \texttt{bp}       & 0.929           & 0.002 & 0.246             & 0.004 \\
                            & & \texttt{d}        & \textbf{0.642}  & 0.005 & \textbf{0.052}    & 0.001 \\
                            & & \texttt{all}      & \textbf{0.720}  & 0.002 & 0.151             & 0.001 \\
        \midrule
        \multirow{14}{*}{\rotatebox[origin=c]{90}{$k = 10$}}
            & \multirow{4}{*}{\shortstack{Perch\\(multiclass)}}
                              & \texttt{bmabz}    & 0.579           & 0.002 & 0.147             & 0.001 \\
                            & & \texttt{bp}       & 0.783           & 0.004 & 0.285             & 0.003 \\
                            & & \texttt{d}        & 0.583           & 0.005 & 0.035             & 0.001 \\
                            & & \texttt{all}      & 0.648           & 0.002 & 0.156             & 0.001 \\
            & {} & {} & {} & {} & {} & {} \\
            & \multirow{4}{*}{\shortstack{Perch + adjustment\\(multiclass)}}
                              & \texttt{bmabz}    & 0.545           & 0.002 & 0.133             & 0.001 \\
                            & & \texttt{bp}       & \textbf{0.950}  & 0.001 & \textbf{0.350}    & 0.005 \\
                            & & \texttt{d}        & 0.622           & 0.004 & 0.046             & 0.001 \\
                            & & \texttt{all}      & 0.706           & 0.002 & \textbf{0.176}    & 0.002 \\
            & {} & {} & {} & {} & {} & {} \\
            & \multirow{4}{*}{\shortstack{Perch + adjustment\\(multilabel)}}
                              & \texttt{bmabz}    & \textbf{0.607}  & 0.002 & \textbf{0.166}    & 0.001 \\
                            & & \texttt{bp}       & 0.943           & 0.001 & 0.295             & 0.004 \\
                            & & \texttt{d}        & \textbf{0.666}  & 0.004 & \textbf{0.059}    & 0.001 \\
                            & & \texttt{all}      & \textbf{0.739}  & 0.002 & 0.174             & 0.002 \\
        \bottomrule
    \end{tabular}
    \label{tbl:final_results_atbfl}
    \vspace{-2.0em}
\end{table}

%%%%%%%%%%%%%%%
% ATBFL k-sweep
%%%%%%%%%%%%%%%

%%%%%%%%%%%
% ATBFL ROC
%%%%%%%%%%%

\vspace{-0.5em}
\subsection{\acs{fb} dataset}

\begin{figure}
    \centering
    \begin{subfigure}[b]{\columnwidth}
        \centering
        \includegraphics[width=\columnwidth]{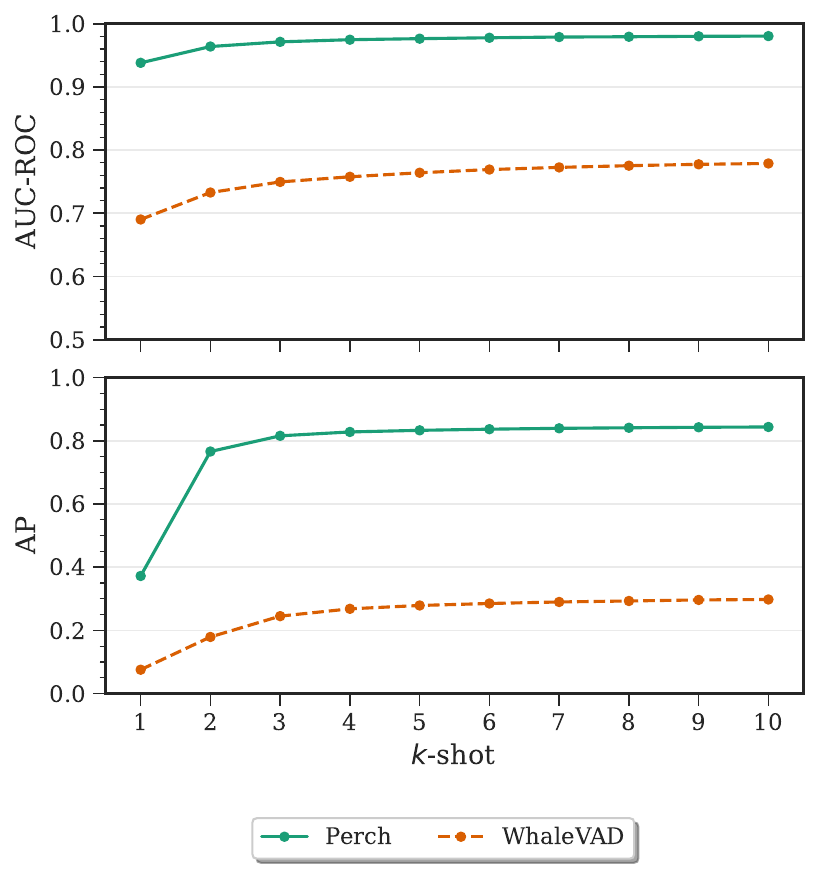}
        \caption{
            Binary detection, in which a segment is positive if it contains any Bryde's whale vocalisation.
        }
        \label{fig:fb_ksweep_detection}
    \end{subfigure}
    \\[1em]
    \begin{subfigure}[b]{\columnwidth}
        \centering
        \includegraphics[width=\columnwidth]{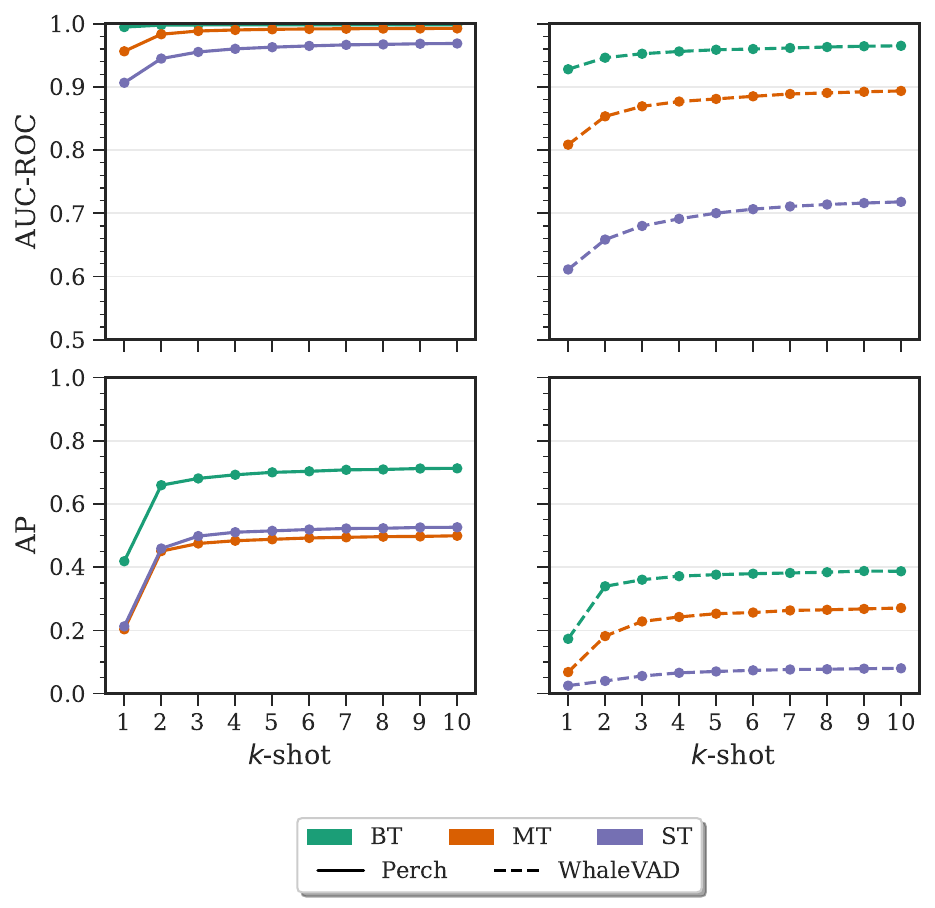}
        \caption{
            Per-call-type classification.
        }
        \label{fig:fb_ksweep_classes}
    \end{subfigure}
    \caption{
        \acs{aucroc} and \acs{ap} on the \acs{fb} dataset as a function of the support set size $k$, comparing \gls{perch} and \gls{wad} as embedding networks.
        No prior score exists, since this is the first time scores are reported on the strongly labelled recording (\texttt{2023-04-22}).
    }
    \label{fig:fb_ksweep}
\end{figure}

\Cref{fig:fb_ksweep_detection} shows the performance achieved by the prototypical classifier in conjunction with the two embedding networks when applied to the \ac{fb} dataset and when the task is reduced to binary detection, where a segment is considered positive if it contains any Bryde's whale vocalisation.
\gls{perch} markedly outperforms \gls{wad} across all values of $k$, achieving an \ac{aucroc} of \num{0.958} and an \ac{ap} of \num{0.372} from a single support example, and \num{0.980} and \num{0.844} at $k = \num{10}$.
The corresponding \gls{wad} scores are \num{0.690} and \num{0.076} at $k = \num{1}$, and \num{0.779} and \num{0.298} at $k = \num{10}$.
Given that annotated calls occupy only \qty{0.38}{\percent} of the queried recording, an \ac{ap} of \num{0.844} represents good separation of call from non-call segments.
% \cmg{@GUNTHER, this \acs{ap} is given as 0.844 in the preceding paragraph and 0.845 here, and the $k=1$ value is 0.372 there and 0.373 below. Neither appears in any table, so I cannot tell which is right. Which is it? }
The sharp increase in \ac{ap} between $k = \num{1}$ and $k = \num{2}$, from \num{0.372} to \num{0.766}, further suggests that a single support segment is frequently insufficient to place the prototype reliably, and that a second example is enough to average out much of this variability.
% \cmg{There is an exact explanation available rather than a suggestion: a prototype is a sample mean, so its variance about the class mean falls as $1/k$, and the largest single reduction available is therefore the halving from the second support example. That explains the shape of every $k$ curve in the paper at once, on both datasets, and is corroborated by the standard deviations falling from 0.024 to 0.007 for \texttt{BT} \acs{ap}. But this would be stealing a bit from my paper.}
% \gt{If it is a bit of stealing then I'd rather leave it as a suggestion I think.}

\Cref{tbl:final_results_fb} and \Cref{fig:fb_ksweep_classes} present the corresponding per-class results.
\gls{perch} outperforms \gls{wad} for every call type at every support set size, with the macro \ac{aucroc} improving from \num{0.859} to \num{0.987} and the macro \ac{ap} from \num{0.246} to \num{0.580} at $k = \num{10}$, absolute improvements of \qty{12.8}{\percent} and \qty{33.4}{\percent} respectively.
The \texttt{BT} call is the most readily identified with both embeddings, with \ac{aucroc} scores of \num{0.998} and \num{0.965} respectively.
This is consistent with its wideband and impulsive character, which sets it apart from the surrounding noise.
% \changed{Both systems find the \texttt{BT} call the easiest to identify, achieving \ac{aucroc} scores of \num{0.998} and \num{0.965} respectively, which are attributed to the wideband and impulsive nature of the call, which distinguishes it clearly from the surrounding noise}{The \texttt{BT} call is the most readily identified with both embeddings, at \ac{aucroc} scores of \num{0.998} and \num{0.965} respectively, which is consistent with its wideband and impulsive character setting it apart from the surrounding noise}.
The two narrowband call types, \texttt{MT} and \texttt{ST}, are more difficult for \gls{wad} to detect, with \acp{aucroc} falling to \num{0.894} and \num{0.718}, respectively. 
\Gls{perch} maintains both scores above \num{0.969}.
% \changed{tonal call types}{narrowband call types, \texttt{MT} and \texttt{ST},}
% \changed{reduces}{falls}

\begin{figure}
    \centering
    \includegraphics[width=0.65\columnwidth]{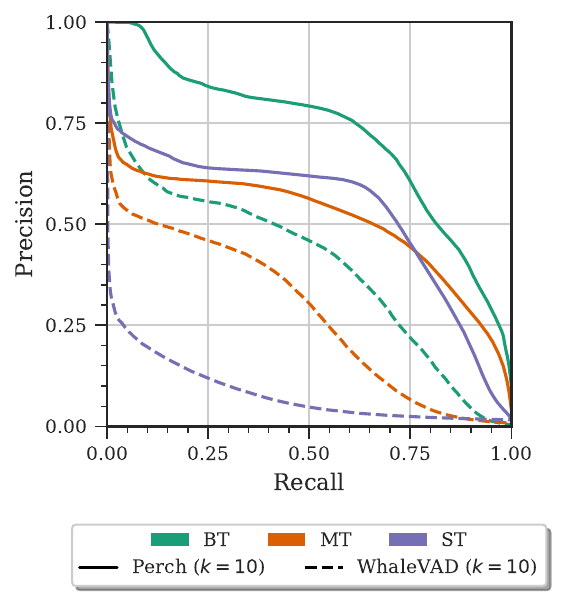}
    \caption{
        Precision-recall curves for each Bryde's whale call type at $k = \num{10}$, comparing \gls{perch} (solid) and \gls{wad} (dashed).
    }
    \label{fig:fb_prcurve}
    \vspace{-2em}
\end{figure}

This result is notable.
\Gls{wad} has been trained on baleen whale vocalisations and might therefore be expected to transfer more readily to an unseen baleen whale species than a model trained predominantly on birdsong.
We attribute this to the task-specific nature of the \gls{wad} representation, which was optimised for the particular structure of blue- and fin-whale calls and therefore generalises poorly to the shorter Bryde's whale calls. %\cmg{@GUNTHER, \textit{shorter} is supported by the duration table, but \textit{higher-frequency} is not supported anywhere: we give the \acs{fb} call frequencies but never the \acs{atbfl} ones, so the comparison cannot be checked. Either add the \acs{atbfl} frequency range, which the front-end argument needs anyway, or drop the word.}
% \remove{highly }
% \changed{, which consequently generalises}{ and therefore generalises}
The much larger and more diverse training set of \gls{perch}, by contrast, appears to yield a representation that is general enough to remain discriminative even for an entirely unseen taxon.
\Cref{fig:fb_prcurve} shows the precision-recall curves obtained at $k = \num{10}$ for each of the three call types.
The \gls{perch} curves lie above those of \gls{wad} over most of the recall range for all three call types, confirming that the improvement is not confined to a particular operating point.
% \cmg{Precision-recall curves over the same evaluation set converge to the class prevalence as recall approaches one, so strict domination \textit{across all values of recall} cannot hold at the right-hand end, and \textit{confirming} overstates what inspecting a figure provides. Suggest \textit{over most of the recall range} and \textit{indicating}.}

%%%%%%%%%%
% FB table
%%%%%%%%%%
\begin{table}
    \centering
    \caption{
        Per-class and macro-averaged (\texttt{all}) \acs{aucroc} and \acs{ap} on the \acl{fb} dataset, for support set sizes $k = \num{1}$, \num{5} and \num{10}.
        Scores are reported as the mean and standard deviation over \num{100} independent support draws.
    }
    \begin{tabular}{@{}ccl|cc|cc@{}}
        \toprule
        {} & \multirow{2}{*}{Model} & \multirow{2}{*}{Type} & \multicolumn{2}{c|}{\acs{aucroc}} & \multicolumn{2}{c}{\acs{ap}} \\
        {} & {} & {} & Avg. & Std. & Avg. & Std. \\
        \midrule
        \multirow{9}{*}{\rotatebox[origin=c]{90}{$k = 1$}}
            & \multirow{4}{*}{\shortstack{Perch + adjustment\\(multilabel)}}
                              & \texttt{BT}       & \textbf{0.995}  & 0.001 & \textbf{0.419}    & 0.024 \\
                            & & \texttt{MT}       & \textbf{0.956}  & 0.004 & \textbf{0.203}    & 0.008 \\
                            & & \texttt{ST}       & \textbf{0.907}  & 0.004 & \textbf{0.213}    & 0.006 \\
                            & & \texttt{all}      & \textbf{0.953}  & 0.002 & \textbf{0.278}    & 0.009 \\
            & {} & {} & {} & {} & {} & {} \\
            & \multirow{4}{*}{\gls{wad}}
                              & \texttt{BT}       & 0.928           & 0.006 & 0.173             & 0.014 \\
                            & & \texttt{MT}       & 0.809           & 0.008 & 0.068             & 0.005 \\
                            & & \texttt{ST}       & 0.611           & 0.008 & 0.025             & 0.001 \\
                            & & \texttt{all}      & 0.783           & 0.004 & 0.089             & 0.005 \\
        \midrule
        \multirow{9}{*}{\rotatebox[origin=c]{90}{$k = 5$}}
            & \multirow{4}{*}{\shortstack{Perch + adjustment\\(multilabel)}}
                              & \texttt{BT}       & \textbf{0.998}  & 0.000 & \textbf{0.700}    & 0.011 \\
                            & & \texttt{MT}       & \textbf{0.991}  & 0.001 & \textbf{0.488}    & 0.008 \\
                            & & \texttt{ST}       & \textbf{0.963}  & 0.002 & \textbf{0.515}    & 0.005 \\
                            & & \texttt{all}      & \textbf{0.984}  & 0.001 & \textbf{0.568}    & 0.004 \\
            & {} & {} & {} & {} & {} & {} \\
            & \multirow{4}{*}{\gls{wad}}
                              & \texttt{BT}       & 0.959           & 0.004 & 0.376             & 0.012 \\
                            & & \texttt{MT}       & 0.881           & 0.005 & 0.253             & 0.009 \\
                            & & \texttt{ST}       & 0.700           & 0.006 & 0.070             & 0.004 \\
                            & & \texttt{all}      & 0.847           & 0.003 & 0.233             & 0.005 \\
        \midrule
        \multirow{9}{*}{\rotatebox[origin=c]{90}{$k = 10$}}
            & \multirow{4}{*}{\shortstack{Perch + adjustment\\(multilabel)}}
                              & \texttt{BT}       & \textbf{0.998}  & 0.000 & \textbf{0.713}    & 0.007 \\
                            & & \texttt{MT}       & \textbf{0.993}  & 0.000 & \textbf{0.500}    & 0.006 \\
                            & & \texttt{ST}       & \textbf{0.969}  & 0.001 & \textbf{0.526}    & 0.005 \\
                            & & \texttt{all}      & \textbf{0.987}  & 0.000 & \textbf{0.580}    & 0.003 \\
            & {} & {} & {} & {} & {} & {} \\
            & \multirow{4}{*}{\gls{wad}}
                              & \texttt{BT}       & 0.965           & 0.003 & 0.387             & 0.010 \\
                            & & \texttt{MT}       & 0.894           & 0.003 & 0.271             & 0.007 \\
                            & & \texttt{ST}       & 0.718           & 0.005 & 0.080             & 0.004 \\
                            & & \texttt{all}      & 0.859           & 0.002 & 0.246             & 0.004 \\
        \bottomrule
    \end{tabular}
    \label{tbl:final_results_fb}
    \vspace{-1.5em}
\end{table}

%%%%%%%%%%%%
% FB k-sweep
%%%%%%%%%%%%

%%%%%%%%%%%%%
% FB PR-Curve
%%%%%%%%%%%%%
\vspace{-0.5em}

%%%%%%%%%%%%%%%%%%%%%%%%%%%%%%%%%%%%%%%%%%%%%%%%%%%%%%
\section{Conclusion}
%%%%%%%%%%%%%%%%%%%%%%%%%%%%%%%%%%%%%%%%%%%%%%%%%%%%%%

In this work, we have investigated whether pretrained bioacoustic embeddings, combined with a prototypical few-shot classifier, can be used to detect and classify baleen whale vocalisations when only a small number of annotated examples are available.
Two pretrained models were compared as embedding networks: \gls{perch}, a general-purpose model trained predominantly on birdsong, and \gls{wad}, a task-specific blue- and fin-whale call detector.
Neither model was fine-tuned, and in all cases classification was performed using at most ten labelled examples per call type.

We found that the default \gls{perch} front-end processing is poorly matched to the low-frequency vocalisations of baleen whales.
The mel filterbank discards the fundamental frequency of most call types and the short frame length affords insufficient frequency resolution in the bands of interest.
By increasing the frame length to \num{8192} samples and restricting the filterbank between \qty{0}{\hertz} and \qty{1000}{\hertz}, the macro \ac{aucroc} and \ac{ap} for \gls{perch} on the \ac{atbfl} dataset at $k=\num{5}$ improved by \qty{5.4}{\percent} and \qty{1.6}{\percent} absolute respectively.
The subsequent introduction of a multilabel one-vs-background formulation, which allows a segment to have more than one label, yielded a further improvement, resulting in a macro \ac{aucroc} of \num{0.739} at $k = \num{10}$.
Taken together, these two modifications represent a \qty{9.1}{\percent} absolute improvement in macro \ac{aucroc} over the unmodified \gls{perch} system at $k = \num{10}$.
% \add{in macro \ac{aucroc} }
% \add{ at $k = \num{10}$}

On the \ac{atbfl} dataset, the few-shot systems do not reach the performance of the fully supervised \gls{wad} topline, which retains a margin of \qty{7}{\percent} in \ac{aucroc} and \qty{19}{\percent} in \ac{ap} at $k = \num{10}$.
This gap is nevertheless modest when considering that \gls{perch} has never been exposed to marine audio and observes at most ten examples per call type, while \gls{wad} was trained on more than \num{58000} annotated blue- and fin-whale calls~\autocite{geldenhuys2025WhaleVAD}.

The FB dataset presents a more equal challenge, since neither of these models has seen any of its data during training.
% The relative ordering of the two embedding networks is reversed on the \ac{fb} dataset, where neither model has seen any training data.\cmg{This one is wrong rather than merely loose, and it is the most attackable sentence in the paper. There is no \acs{atbfl} ordering of the two \textit{embedding networks} to reverse: on \acs{atbfl}, \gls{wad} appears only as a fully supervised per-frame topline and is never used as an embedding network. Suggest \textit{The FB dataset is the only one on which the two models are compared in the same role, and neither has seen any of its training data.}}
In this case \gls{perch} outperforms \gls{wad} for every call type and at every support set size, achieving a macro \ac{aucroc} of \num{0.987} and a macro \ac{ap} of \num{0.580} at $k = \num{10}$, against \num{0.859} and \num{0.246} for \gls{wad} respectively.
Given a pretrained embedding, it therefore appears that a handful of examples from an unseen target species outweigh a much larger and painstakingly labelled dataset of related species.
% For an unseen species, the diversity of the pre-training dataset therefore appears to outweigh the similarity in target domain.
% We attribute this to the task-specific nature of the \gls{wad} representation, which was optimised for the particular structure of blue- and fin-whale calls and consequently transfers poorly to the shorter and higher-frequency Bryde's whale vocalisations.\cmg{These two sentences repeat the results section almost verbatim.}
% The broader and more diverse training set of \gls{perch}, by contrast, appears to yield a representation that remains discriminative even for an entirely unseen taxon.

These results suggest that a general-purpose bioacoustic embedding, paired with a lightweight few-shot classifier, offers a practical and performant starting point for the discovery and monitoring of call types for which annotated data is scarce or absent.
% \remove{, and that the front-end configuration of such a model warrants careful attention when it is applied outside the frequency range for which it was designed}
They also show that the input representation of such a model requires attention whenever it is applied to frequency content for which it was not designed.
% \add{They also show that the input representation of such a model requires attention whenever it is applied outside the frequency range for which it was designed.}
Future work will consider the adaptation of the embedding network itself to marine audio, the extension of the approach from segment-level classification to frame-level event detection, and the inclusion of call types such as the tonal downsweep, which could not be evaluated here owing to its occurrence in only a single recording.

\vspace{-0.35em}
\section*{Acknowledgements}
\noindent
\blindpar{%
The authors gratefully acknowledge Telkom (South Africa) for their financial support and thank the Stellenbosch Rhasatsha high performance computing~(HPC1) facility for compute time provided to the research presented in this paper.
The authors also acknowledge the anonymous reviewers for their insightful comments, which improved the clarity of this manuscript.%
}{%
% \vspace{-1.2em}
\small\ttfamily
Acknowledgements have been removed\\
for the purposes of blind review.%
}

\vspace{-0.5em}

\printbibliography
\end{document}